\documentclass[manuscript,screen]{acmart}
\usepackage{natbib}
\AtBeginDocument{%
  \providecommand\BibTeX{{%
    \normalfont B\kern-0.5em{\scshape i\kern-0.25em b}\kern-0.8em\TeX}}}

\setcopyright{acmcopyright}
\copyrightyear{2023}
\acmYear{2023}
\acmDOI{XXXXXXX.XXXXXXX}

\acmConference[CHI '23]{Make sure to enter the correct
  conference title from your rights confirmation emai}{April 23–28,
  2023}{Woodstock, NY}
\acmBooktitle{the workshop "Beyond prototyping boards: future paradigms for electronics toolkits" at the CHI Conference on Human Factors in Computing
Systems (CHI ’23), April 23–28, 2023, Hamburg, Germany. ACM, New York, NY, USA}

\acmPrice{15.00}
\acmISBN{978-1-4503-XXXX-X/18/06}

\begin{document}

\title{Frankenstein's Toolkit: Prototyping Electronics Using Consumer Products}

\author{Ilan Mandel}
\email{im334@cornell.edu}
\affiliation{%
  \institution{Cornell Tech}
  \streetaddress{11 E Loop Road}
  \city{New York}
  \state{NY}
  \country{USA}
  \postcode{10044}
}
\author{Wendy Ju}
\email{wendy.ju@cornell.edu}
\affiliation{%
  \institution{Cornell Tech}
  \streetaddress{11 E Loop Road}
  \city{New York}
  \state{NY}
  \country{USA}
  \postcode{10044}
}




\begin{abstract}
 In our practice as educators, researchers and designers we have found that centering reverse engineering and reuse has pedagogical, environmental, and economic benefits. Design decisions in the development of new hardware tool-kits should consider how we can use e-waste at hand as integral components of electronics prototyping. Dissection, extraction and modification can give insights into how things are made at scale. Simultaneously, it can enable prototypes that have greater fidelity or functionality than would otherwise be cost-effective to produce.
\end{abstract}




\maketitle

\section{Introduction}
In Mary Shelley's 1818 Gothic novel \textit{Frankenstein; or, The Modern Prometheus}~\cite{frankenstein}, the monster is often been depicted as a collage of body parts~\cite{heffernan1997looking}. Just as Victor Frankenstein imbues leftover remains with a spark of life, designers often reanimate and reuse leftover electronics to make new systems. Reuse allows designers to better understand how the devices around them work and get made. Simultaneously, there are economic and ecological benefits for researchers and designers to apply approaches that ask how existing consumer products can be leveraged as a malleable design material. Thus, tools for developing interactive prototypes should enable probing, hacking and reintegration of e-waste and electronics at hand. 
We observe that Integrated Microcontroller Development Boards (see Type 2a) and  Breakout Boards and Wireless Modules (see Type 2b) from Lambrichts et al.'s electronics toolkit taxonomy ~\cite{lambrichts2021survey} are often employed as part of Frankenstein practices. These practices could be further supported if tool-kits provided designers the tools they need to reverse engineer systems, identify electronics components and to parrot signals and states from those components. 

\section{Designing with waste and garbage}
Sociologist Jennifer Gabrys has championed salvage as a sustainable practice wherein waste is transformed into a material resource ~\cite{gabrys2012salvage}. Ethnographers and sociologists have highlighted inspirational salvage and material re-use practices such as ``Jugaad''~\cite{singh2021upcycling} in South Asia and ``Bruck'' in the Orkney islands ~\cite{watts2019energy}. Within these resource-constrained communities and contexts, people apply ingenuity to make new products from old material. It has long been implied that computing devices do not lend themselves well to salvage. Gabrys, for instance, states that "electronics do not simply dissolve into raw materials ready to take on a second life." ~\citet{dew2019designing} created a workshop to engage designers in a reflective process of ecological inversion to sensitize them to material flows; in response, designers indicated a concern that ``[salvaging materials] \textit{made everything harder}'' for their practice. 

However hacking, probing, reverse engineering and salvage have long been part of  electronics maker practices. We find examples of these practices both in academia ~\cite{cao2014enhancing, adams2020puffpacket, hwu2018hacking, trivedi2017android} and by numerous makers who share projects on websites like \href{www.Hackaday.io}{Hackaday}, \href{www.Instructables.com}{Instructables} and \href{www.Github.com}{Github}. These practices involve disassembly to extract components, reprogramming Micro-controllers for added functionality and probing internal signals in closed systems for novel applications. This work often involves learning how existing systems work in order to ``replace their brains'' and re-animate them for the designers purposes. 

As we consider future prototyping hardware tool-kits, we can make existing consumer electronics more malleable, adaptable and sustainable by centering the tools, techniques and pedagogy's of Frankenstein-ing electronics. This work is informed by our experiences using hacked and reused consumer products in a graduate level mobile robotics and interactive device design courses, as well as our own research practice, leveraging these devices to build novel systems.











\section{Consumer Goods as a Prototyping Platform}

A case study can be made of \href{https://tasmota.github.io/docs/}{Tasmota}, an open source firmware for existing IoT devices powered by an ESP8266 micro-controller board. Tasmota was first introduced by Theo Arends in January 2017 to enable SmartHome devices to bypass proprietary cloud control services. The resulting ecosystem allows cheap consumer electronics to function similarly to a Type 3 (Integrated Modular Systems) electronics tool-kit~\cite{lambrichts2021survey}. There are numerous examples of electronics prototyping kits being used to add logging, interactivity, controls, or other``smart'' features to ``dumb'' consumer products. Tasmota instead starts with plentiful, cheap ``smart'' consumer electronics but adds functional controls for end users.

The ESP8266 itself was originally produced by Expressif Systems for home appliances and sensor networks~\cite{mehta2015esp8266}. A cheap development board from AI-Thinker began exciting members of the maker community in 2014~\cite{Benchoff_2014}. At first, the platform was poorly documented and required a dedicated tool-chain. However, a community of developers translated Chinese documentation~\cite{Baguley_2015}, developed new tool-chains and integrated the ESP8266 with the Arduino IDE to make the platform more widely accessible~\cite{Benchoff_2015}. By 2015, makers began finding cheap consumer hardware from Shenzhen ~\cite{lindtner2020prototype} on Aliexpress with ESP8266's inside that they would open and flash custom firmware to~\cite{Benchoff_consumer}. 

Tasmota began as a project to flash one particular ESP8266 based IoT smart switch with a custom firmware to enable MQTT integration and over-the-air updates. The Tasmota platform quickly grew into a fully fledged ecosystem that supports nearly any ESP8266-based consumer good. There are over 2500 templates on a community maintained database that can be easily integrated into a system and programmed with low or no-code interfaces~\cite{blakadder}. These include the gamut of ``smart'' consumer electronics including lights, speakers, switches, outlets, air quality, temperature and other sensors. There are templates for standard ESP8266 development boards for more flexible additions to a Tasmota-based system. We argue that the Tasmota functions as an electronics prototyping tool-kit whose modules are comprised of cheap networked consumer electronics. Students in our lab use Tasmota-flashed smart plugs to prototype room-scale interactive devices. This "hack" gives students the ability to develop systems that required switching 110V AC power using 5v DC electronics. 

\section{Reverse Engineering in Electronics Design Pedagogy}
In our interactive device design courses, we have extended introductory programming exercises such as Blink to ``FrankenBlink''.  After learning to ``blink'' the onboard LED, students are asked to blink an LED on an old, broken or unused consumer electronic device or use the input devices such as buttons and dials. 

~\citet{sheppard1992mechanical} describes the pedagogical value of teaching students to answer ``how did others solve this problem'' via mechanical dissection and disassembly. This model has successfully been applied to electronics ~\cite{rouse2022transforming, sandborn2006using}.~\citet{lambrichts2021survey}  discuss the challenges of moving from prototypes to small multiples, scaling up and manufacturing.  We believe that knowledge gained through disassembly and hacking can help students understand how design for manufacturing and scale operate. 
In doing so, we can give designers better tools for taking concepts from prototypes to products. 

Web design and front-end development are a useful analog to understand how people learn from dissecting systems. The ability to view a web page's source code~\cite{o2009web, shirky1998and}, or copy and remix elements on websites such as MySpace ~\cite{miltner2022tom} serve as the nostalgic origin story for countless web-developers. Tooling such for web inspection such as Firebug~\cite{luthra2010firebug}, which has been eventually integrated into every major browser, gave users new tools for modifying, editing and programming on top of the web~\cite{tanner2019poirot}. We believe prototyping kits can act as dissection tools for physical systems, helping users understand how consumer products are put together so the parts of these products can be edited, emulated and reused.

The design of future electronics tool-kits should encourage users to tap into the signals and systems of electronics they already interact with. This consideration would give new groups of hardware developers the ability to ``right click and view source'' on the products around them.




\section{Resilient Prototyping, Sustainability and Frugality}

Electronics prototyping exists within a complex supply chain of production that is not isolated from resource and geopolitical constraints. In the midst of Covid-19 and unprecedented chip shortages, the Raspberry Pi 4 is being sold at a 400\% mark up~\cite{Pounder_2022} which made it too expensive for student coursework. In our case, we scavenged Raspberry Pi 3's for students. However, hackers and reverse engineers have long managed to use old routers~\cite{Eliot_2005}, TV-set top boxes~\cite{Devreker} and other kinds of e-waste as cheap Linux machines with similar functionality to a Raspberry Pi. We developed a mobile robotics platform made from re-used hoverboard components that is both more powerful and less expensive than existing commercial options. We have deployed this platform in our research and a graduate level mobile robotics course. Reuse can provide solutions during acute shortages, while also reducing long-term ecological impact.

The core processes of making and rapid prototyping present inherent unresolved sustainability questions ~\cite{arroyos2022tale, roedl2015sustainable, lazaro2020introducing}. Centering Frankenstein-ing electronics in electronics prototyping tool-kits incentivizes methods that leverage e-waste, and emphasizes adaptive reuse~\cite{singh2021upcycling} and repair~\cite{houston2016values}. 



\section{Discussion}

We see promising trends in the developments of electronics prototyping systems. Programmable digital peripheral components such as the Programmable Input/Output (PIO) on the RP2040 enable developers to interface with old, obsolete, or obscure protocols and hardware ~\cite{smith2022initialize}. The PIO pins also enable cheap development boards to function as a logic analyzers~\cite{alves2016using} or oscilloscopes~\cite{Hill_2022}. Open source tooling and hardware such as ChipWhisperer~\cite{o2014chipwhisperer} are making more advanced reverse engineering techniques such as side channel attacks, power analysis attack, voltage and clock glitching more accessible to new cohorts of engineers and designers. 

The physical affordances for electronic prototyping tools could improve users ability to identify and reuse components in existing products. The Superprobe implemented by Dangerous Prototypes takes the form factor of a handheld probe and acts as an advanced multi-meter and parts analyzer~\cite{Ian_2011}. The Flipper Zero is multi-tool for pen-testers. It has a wide range tools for interacting with  radio protocols for wireless devices and GPIO's for interfacing with hardware~\cite{Person_2022}. Design features such as a portable gameified form factor along with and tamogochi-like dolphin character introduce new users to hacking, reverse engineering and modification of electronics.

We acknowledge that an emphasis on Frankenstein-ing poses additional challenges for the rapid development of electronic prototypes. However we suggest that there are often economic, pedagogic or sustainability arguments for making these trade offs. As we consider what forms and primitives future electronics tool-kits might take, we hope designers will iterate on the state of the art while keeping these approaches in mind.




\bibliographystyle{ACM-Reference-Format}
\bibliography{bibliography}

\end{document}